\PassOptionsToPackage{table, xcdraw}{xcolor}

\documentclass[acmsmall]{acmart}

\usepackage[normalem]{ulem}
\usepackage{tabularray}
\usepackage{graphicx}
\usepackage{colortbl}
\AtBeginDocument{%
  }

\setcopyright{cc}
\setcctype{by}
\acmJournal{PACMHCI}
\acmYear{2025} \acmVolume{9} \acmNumber{7} \acmArticle{CSCW368} \acmMonth{11} \acmPrice{}\acmDOI{10.1145/3757549}

\acmISBN{978-1-4503-XXXX-X/18/06}

\begin{document}

\title[AI That Helps Us Help Each Other: A Proactive System for Scaffolding \\ Mentor-Novice Collaboration in Entrepreneurship Coaching]{AI That Helps Us Help Each Other: A Proactive System for Scaffolding Mentor-Novice Collaboration in Entrepreneurship Coaching}

\author{Evey Jiaxin Huang}
\email{evey.huang@kellogg.northwestern.edu}
\orcid{0000-0001-7092-0898}
\affiliation{%
  \institution{Northwestern University}
  \city{Evanston}
  \state{Illinois}
  \country{USA}
}

\author{Matthew Easterday}
\email{easterday@northwestern.edu}
\orcid{0000-0002-0101-7440}
\affiliation{%
  \institution{Northwestern University}
  \city{Evanston}
  \state{Illinois}
  \country{USA}
}

\author{Elizabeth Gerber}
\email{egerber@northwestern.edu}
\orcid{0000-0001-8914-071X}
\affiliation{%
  \institution{Northwestern University}
  \city{Evanston}
  \state{Illinois}
  \country{USA}
}
\renewcommand{\shortauthors}{Evey Jiaxin Huang, Matthew Easterday, and Elizabeth Gerber}

\begin{abstract}
   Entrepreneurship requires navigating open-ended, ill-defined problems: identifying risks, challenging assumptions, and making strategic decisions under deep uncertainty. Novice founders often struggle with these metacognitive demands, while mentors face limited time and visibility to provide tailored support. We present a human-AI coaching system that combines a domain-specific cognitive model of entrepreneurial risk with a large language model (LLM) to proactively scaffold both novice and mentor thinking. The system proactively poses diagnostic questions that challenge novices’ thinking and helps both novices and mentors plan for more focused and emotionally attuned meetings. Critically, mentors can inspect and modify the underlying cognitive model, shaping the logic of the system to reflect their evolving needs. Through an exploratory field deployment, we found that using the system supported novice metacognition, helped mentors plan emotionally attuned strategies, and improved meeting depth, intentionality, and focus---while also surfaced key tensions around trust, misdiagnosis, and expectations of AI. We contribute design principles for proactive AI systems that scaffold metacognition and human-human collaboration in complex, ill-defined domains, offering implications for similar domains like healthcare, education, and knowledge work. 
\end{abstract}



\received{October 2024}
\received[revised]{April 2025}
\received[accepted]{August 2025}


\begin{teaserfigure}
\centering
\includegraphics[width=\textwidth]{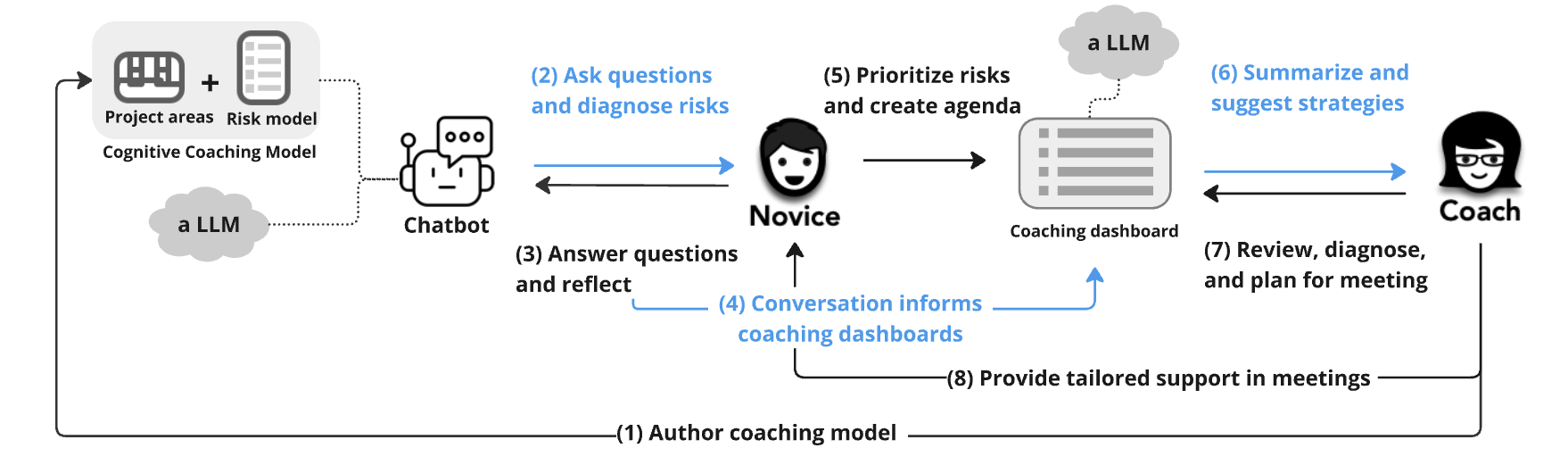}
\caption{The human-AI coaching system leverages a cognitive coaching model and LLM to proactively scaffold novice reflection, support mentor preparation, and adapt to evolving contexts.}
\label{fig:s3-flow}
\end{teaserfigure}

\maketitle

\section{Introduction}
To design novel and effective solutions that tackle real-world, complex problems such as providing equitable healthcare and reducing the impact of global warming, entrepreneurs must identify what problem to solve, who their users are, and whether their assumptions are valid---all in the absence of clear goals, constraints, or evaluation criteria~\cite{jonassenAllProblemsAre2008, lynchDefiningIllDefinedDomains}. These challenges are not just technical; they are cognitive and emotional, requiring entrepreneurs to reason through ambiguity, reflect on evolving priorities, and confront risks that may cause their solutions to fail~\cite{dorstCreativityDesignProcess2001, jonassenDesignTheoryProblem2000}. Novice entrepreneurs, particularly those who are first-time founders, often struggle in this terrain, not because they lack intelligence or commitment, but because they are still developing metacognitive skills, such as risk diagnosis, articulation, and planning that experienced entrepreneurs rely on~\cite{schonEducatingReflectivePractitioner1987, ackermanMetaReasoningMonitoringControl2017, adamsEducatingEffectiveEngineering2003, reiserScaffoldingComplexLearning2004, huangIntelligentCoachingSystems2023}. As a result, novices frequently focus on surface-level tasks, overlook critical risks, and struggle to turn uncertainty into strategic insight~\cite{krajcikInquiryProjectBasedScience1998, chiCategorizationRepresentationPhysics1981}. 

Coaching plays a pivotal role in supporting novice entrepreneurs to tackle real-world, ill-defined problems~\cite{jonassenAllProblemsAre2008,reeslewisOpportunitiesEducationalInnovations2019, lynchDefiningIllDefinedDomains, huangIntelligentCoachingSystems2023}. Mentors, often former founders or practitioners, coach novices not by prescribing answers but by cultivating metacognitive practices: diagnosing risks, interrogating assumptions, and preparing them to act despite uncertainty~\cite{pittawayEntrepreneurshipEducationSystematic2007, reeslewisPlanningIterateSupporting2018, reeslewisBuildingSupportTools2015, collinsCognitiveApprenticeship2005, luProblembasedLearning2014}. However, mentors are often stretched thin, coaching multiple novice teams simultaneously across diverse and evolving problem spaces~\cite{huangIntelligentCoachingSystems2023, reeslewisBuildingSupportTools2015, gargOrchestrationScriptsSystem2023}. They must interpret sparse updates, diagnose risks and root causes, and plan targeted assistance---all under time constraints~\cite{reeslewisOpportunitiesEducationalInnovations2019, crossExpertiseDesignOverview2004, carlsonDesignRisksFramework2020, collinsCognitiveApprenticeshipTeaching1989, riesLeanStartupHow2017, huangIntelligentCoachingSystems2023}. Existing support tools such as templates and task trackers offer structure but require mentors to do the interpretive heavy lifting and are poorly suited for ill-defined, real-time collaboration~\cite{zhangAgileResearchStudios2017, gargOrchestrationScriptsSystem2023, simonStructureIllStructured1973, koedingerCognitiveTutorsTechnology2006}. 

Recent advances in LLMs open possibilities for more adaptive support in open-ended domains~\cite{bubeckSparksArtificialGeneral2023}. However, most LLM-based tools remain passive: they follow the user’s framing, offer generic advice, and rarely challenge reasoning~\cite{sarkarAIShouldChallenge2024}. Even reflective systems like \textit{Thinking Assistants}~\cite{parkThinkingAssistantsLLMBased2024} and \textit{Rehearsal}~\cite{shaikhRehearsalSimulatingConflict2024} focus on individual users, rely on static expert knowledge, and operate in simulated problems. These limitations make them ill-suited for coaching entrepreneurs, which involves asymmetrical roles, dynamic goals, and real-world problem solving~\cite{huangIntelligentCoachingSystems2023, reeslewisAssessingIterativePlanning2019, gargOrchestrationScriptsSystem2023}. We argue that supporting entrepreneurship coaching requires a new class of AI systems: ones that are proactive, grounded in domain-specific expertise, capable of adapting over time, and designed to scaffold multi-user collaboration in live, high-uncertainty problem contexts---augmenting, not replacing, human-human interaction.

In this paper, we explore how AI systems might proactively scaffold metacognitive processes in entrepreneurship coaching---such as risk diagnosis, reflection, and strategy planning---while also supporting the judgment and intentionality of human mentors. To guide this investigation, we ask: 
\begin{enumerate} 
    \item How can an AI system scaffold the metacognitive processes of novice entrepreneurs, such as articulating updates, identifying critical risks, and planning the next steps? 
    \item In what ways can such a system support mentors in interpreting novice thinking, identifying root causes, and preparing targeted, context-sensitive coaching strategies? \item How does integrating this system shape the focus, depth, and efficiency of mentor–novice coaching conversations? 
\end{enumerate}

Through a Research through Design (RtD) approach~\cite{zimmermanResearchDesignMethod2007b}, we designed and developed a coaching system that integrates cognitive coaching models with LLMs to scaffold both novices and mentors through structured reflection and preparation. The system proactively identifies potential risks based on novices' inputs and structures them on a dashboard for mentors, who can also inspect and adapt the logic of the system's reasoning. We deployed this system in a university incubator, where one mentor and eleven novice entrepreneurs used it to prepare for and engage in real coaching meetings. Our findings show that the system helped novices uncover blind spots, clarify priorities, and engage more deeply with reflection. Mentors reported increased intentionality, reduced cognitive burden, and more focused, emotionally attuned conversations. 

This work contributes to human–AI collaboration, CSCW, and the Learning Sciences by advancing how we design AI systems to scaffold reflection and collaboration in ill-defined domains like entrepreneurship. First, we contribute to human–AI collaboration by introducing a system architecture that integrates expert-governed, cognitive coaching models with LLMs to support proactive, inspectable, and domain-specific reasoning---shifting from task automation to scaffolding diagnostic and relational work. We derive five design principles for such systems: (1) proactively challenge novices’ thinking on risks, (2) layer dual context from both mentor and novice to adapt support, (3) empower mentors to inspect and shape domain-grounded AI logic, (4) surface root causes---both cognitive and emotional---to scaffold mentor insight, and (5) orchestrate human-human collaboration through asynchronous, role-sensitive preparation. Second, we contribute to the Learning Sciences by articulating how these principles enable AI systems to support metacognitive skills like risk diagnosis and strategic planning through context-aware, reflective prompts that preserve user agency. Finally, we contribute to CSCW by providing empirical insights into how AI can enhance---not replace---human-human collaboration in ill-defined domains, offering a concrete implementation of AI as infrastructure for intentional, emotionally attuned, and cognitively focused mentorship.

\section{Background}

\subsection{Entrepreneurship Coaching}
Entrepreneurs need to design innovative and effective solutions to tackle societal challenges that are complex and ill-defined~\cite{jonassenEverydayProblemSolving2006}, lacking predefined solution paths or singular correct answers~\cite{jonassenAllProblemsAre2008, lynchDefiningIllDefinedDomains}. At the outset of a project, many subproblems and design elements remain unknown, requiring entrepreneurs to continuously discover and define them throughout the design process~\cite{dorstCreativityDesignProcess2001, jonassenDesignTheoryProblem2000}. Researchers have identified meta-cognition as a key driver in the successful resolution of poorly defined problems~\cite{schonEducatingReflectivePractitioner1987, ballAdvancingUnderstandingDesign2019, winneMetacognition2014, ackermanMetaReasoningMonitoringControl2017}. Meta-cognition involves cognitive processes that monitor, reflect on, and enhance one's problem-solving strategies~\cite{brownMotivationLearnUnderstand1988, flavellMetacognitionCognitiveMonitoring1979, flavellMetacognitiveAspectsProblem1976}. However, novice entrepreneurs, especially those who are first-time founders~\cite{huangIntelligentCoachingSystems2023}, have not yet developed effective  metacognitive skills such as monitoring progress, diagnosing potential risks that may cause their design to fail, and planning actions to address these critical risks~\cite{bouwmanHumanDiagnosticReasoning1983, johnsonCognitiveAnalysisExpert1988, koedingerExploringAssistanceDilemma2007, reiserScaffoldingComplexLearning2004, carlsonDesignRisksFramework2020}.

Coaching is one of the most effective methods to help novice entrepreneurs develop the metacognitive skills necessary to solve ill-defined, real-world problems~\cite{collinsCognitiveApprenticeshipTeaching1989, hmelo-silverGoalsStrategiesProblembased2006, hmelo-silverProblemBasedLearningWhat2004, luProblembasedLearning2014}. Coaching is prevalent in problem-based learning~\cite{hmelo-silverGoalsStrategiesProblembased2006, hmelo-silverFacilitatingCollaborativeKnowledge2008}, project-based learning~\cite{gilbuenaFeedbackProfessionalSkills2015, zhangAgileResearchStudios2017}, and design education~\cite{reeslewisPlanningIterateSupporting2018} because it is adaptive and enables novices to learn expert-like strategies for tackling ill-defined problems~\cite{collinsCognitiveApprenticeshipTeaching1989}. Mentors monitor novices' progress, diagnose challenges and potential risks that could lead to project failure, and help them adopt effective strategies to address these risks~\cite{collinsCognitiveApprenticeshipTeaching1989, gilbuenaFeedbackProfessionalSkills2015, huangIntelligentCoachingSystems2023}. This hands-on support is more effective than less personalized approaches such as workshops and lectures, which may be too abstract or not immediately applicable~\cite{langleyImprovementGuidePractical2009, princeInductiveTeachingLearning2006, darling-hammondTeacherEducationWorld2017}.

Although effective, coaching is labor-intensive and cognitively demanding. Unlike in well-defined domains like algebra, mentors cannot simply check whether novices have reached the correct solutions (which do not exist for ill-defined problems~\cite{lynchDefiningIllDefinedDomains, jonassenAllProblemsAre2008}); instead, they must assess the reasonableness of proposed designs given novices' current understanding of the problem~\cite{carlsonDesignRisksFramework2020, carlsonDefiningAssessingRisk2018} and how novices are working to gain that understanding ~\cite{gargOrchestrationScriptsSystem2023, zhangAgileResearchStudios2017}. To do so, mentors must spend considerable time and labor to acquire a wide range of often invisible information, including, but not limited to: (1) the understanding of the problem and rationale for their design choices~\cite{reeslewisBuildingSupportTools2015, crossExpertiseDesignOverview2004, atmanVerbalProtocolAnalysis1998}; (2) how novices are working, what they have done and plan to do~\cite{gargOrchestrationScriptsSystem2023, zhangAgileResearchStudios2017, huangIntelligentCoachingSystems2023}; and (3) novices' emotions---whether they feel confident, motivated or discouraged~\cite{gorsonUsingElectrodermalActivity2022, safranEmotionPsychotherapyChange1991, greenbergVarietiesShameExperience1997}. Even if all this information becomes visible, the sheer volume and complexity of the information can be overwhelming and cognitively demanding for mentors to diagnose potential risks and root causes~\cite{carlsonDefiningAssessingRisk2018, klahrCognitiveObjectivesLOGO1988}. Furthermore, mentors need to modulate their strategies to balance providing effective guidance while fostering novices' independent problem-solving abilities ~\cite{koedingerExploringAssistanceDilemma2007, lepperMotivationalTechniquesExpert1993, merrillTutoringGuidedLearning1995}. These challenges become exponentially harder as the number of novices increases---especially when each is working on a different project, at a different pace, and with unique contextual needs~\cite{huangIntelligentCoachingSystems2023}. 

\subsection{Existing Approaches to Support Coaching}
A longstanding area of HCI and CSCW research has explored how technology can support coaching by scaffolding mentor-novice interactions. One widely adopted strategy is the use of structured templates, such as Lean Canvas or storyboards, to scaffold novices to articulate elements of their project such as value propositions, customer segments, and risks~\cite{riesLeanStartupHow2017, reeslewisBuildingSupportTools2015}. These tools externalize novice thinking and make it legible to mentors, reducing the time that mentors must spend probing basic assumptions
~\cite{reeslewisBuildingSupportTools2015, reeslewisAssessingIterativePlanning2019, gargOrchestrationScriptsSystem2023}. Yet, the rigidity of templates may constrain creative thinking and fail to capture the complexity and fluidity of real-world entrepreneurial challenges~\cite{macneilProbMapAutomaticallyConstructing2021}. Task tracking and documentation tools (for example, Asana, Kaleidoscope~\cite{stermanKaleidoscopeReflectiveDocumentation2023}) have also been used to show the progress of the work of novices and planning processes~\cite{stermanKaleidoscopeReflectiveDocumentation2023, gargOrchestrationScriptsSystem2023}. However, these tools depend on the consistent and high-quality engagement of novices - something that is often lacking due to limited motivation or perceived value~\cite{moonReflectionLearningProfessional2000, reeslewisBuildingSupportTools2015, stermanKaleidoscopeReflectiveDocumentation2023}. Furthermore, these tools still require mentors to review and interpret fragmented data across tools and sessions, which is cognitively demanding and does not scale when mentors support multiple novice teams~\cite{boudReflectionTurningExperience1987}.

Another stream of work has examined how peer mentoring and collaborative learning models can support novice development by distributing the mentoring load~\cite{kulkarniPeerStudioRapidPeer2015, shannonPeerPresentsWebBasedSystem2016, xuVoyantGeneratingStructured2014}. Peer coaching leverages novices' collective knowledge and experiences by facilitating collaboration and mutual support. Through group discussions, feedback sessions, and collaborative projects, novices learn from each other under the guidance of a mentor~\cite{zhangAgileResearchStudios2017, hadwinSelfRegulationCoRegulationShared2017, laveSituatedLearningLegitimate1991}. Although effective in many learning domains, peer-based approaches face critical limitations in entrepreneurship coaching. Novices often work on different ventures and progress at varying paces, making it difficult to identify others who possess relevant knowledge to support them~\cite{huangIntelligentCoachingSystems2023}. The uniqueness and diversity of each entrepreneurial project mean that the experiences and insights of a novice may not be directly applicable to the situation of another~\cite{pittawayEntrepreneurshipEducationSystematic2007, copeLearningDoingExploration2000}. Furthermore, mentors may find it challenging to monitor all interactions, potentially missing opportunities to correct misunderstandings~\cite{banduraSocialLearningTheory1977}.

Other systems support mentors by offering decision support through real-time analytics. For example, classroom dashboards have been co-designed with teachers to present actionable insights based on student real-time engagement and performance data~\cite{holsteinClassroomDashboardCodesigning2018}. Many of these systems are based on AI's predictive capabilities and are trained on large-scale, structured datasets~\cite{caiHelloAIUncovering2019, burgessHealthcareAITreatment2023, patelHumanMachinePartnership2019, liuChipNeMoDomainAdaptedLLMs2024, cabreraImprovingHumanAICollaboration2023}. Such systems have shown value in domains such as healthcare and education, where well-defined data structures and task outcomes support algorithmic inference~\cite{jacobsDesigningAITrust2021c, caiHelloAIUncovering2019, holsteinCoDesigningRealTimeClassroom2019}. However, entrepreneurship is fundamentally different: It is a domain of ill-defined, evolving problems with no fixed success metrics, no labeled training data, and no ground truth against which novice thinking can be benchmarked~\cite{simonStructureIllStructured1973, jonassenDesignTheoryProblem2000, dorstCreativityDesignProcess2001}. Even more critically, the most essential input to mentor decision-making - such as the reasoning, assumptions, and internal roadblocks of novices - are rarely visible, often unarticulated, and rarely captured in existing tools or datasets~\cite{kotteEntrepreneurialCoachingTwoDimensional2021, carlsonDesignRisksFramework2020, reeslewisAssessingIterativePlanning2019}.

\subsection{AI Approaches to Automate Coaching}
In parallel, HCI and AI researchers have explored how AI systems can automate or augment coaching and feedback processes. Intelligent Tutoring Systems (ITS) are one such approach, using structured cognitive models and encoded solution paths to provide personalized, real-time guidance~\cite{butzWebBasedIntelligentTutoring2004, koedingerNewPotentialsDataDriven2013, vanlehnAndesPhysicsTutoring2005, virvouCollaborativeSupportMultilingual2012}. These systems are highly effective in well-defined domains such as algebra where task structures are known and expert reasoning can be modeled explicitly~\cite{vanlehnBehaviorTutoringSystems2006a, koedingerCognitiveTutorsTechnology2006, redfieldFutureIntelligentTutoring1991}. However, entrepreneurship coaching poses a fundamentally different design space: novices face ill-defined, open-ended problems where goals, success criteria, and solution paths are often ambiguous~\cite{jonassenInstructionalDesignModels1997, simonStructureIllStructured1973, jonassenDesignTheoryProblem2000}. This makes it difficult, if not impossible, to pre-define correct answers or construct comprehensive knowledge bases for comparison~\cite{ackerGeneratingCoherentExplanations1991, redfieldFutureIntelligentTutoring1991}. 

More recently, LLMs have enabled AI systems to have new possibilities to reason and communicate in open domains~\cite{bubeckSparksArtificialGeneral2023}. LLM-powered tools have been deployed to summarize meetings~\cite{noyExperimentalEvidenceProductivity2023a}, generate writing assistance~\cite{biermannToolCompanionStorywriters2022}, and support personalized instruction in domains such as essay writing and algebra~\cite{daiCanLargeLanguage2023, tackAITeacherTest2022}. However, most of these systems act as reactive assistants---responding to user inputs without proactively identifying gaps in reasoning, facilitating multi-party collaboration, or supporting metacognitive development~\cite{sarkarAIShouldChallenge2024, sarkarExploringPerspectivesImpact2023, tankelevitchMetacognitiveDemandsOpportunities2023, gmeinerExploringChallengesOpportunities2023}.

Recent work has explored how LLMs can scaffold reflection and reasoning in open-ended domains~\cite{parkThinkingAssistantsLLMBased2024, shaikhRehearsalSimulatingConflict2024}. For example, \textit{Thinking Assistants} by Park et al.~\cite{parkThinkingAssistantsLLMBased2024} leverage LLMs to support productive self-reflection by asking questions rather than offering answers. However, this system is designed for individual use (e.g., exploring career decisions) and replaces mentors completely. They fall short of addressing the relational and role-asymmetric dynamics of mentorship~\cite{huangIntelligentCoachingSystems2023}. In addition, expert knowledge in that system is static: experts can verify embedded content but cannot modify or adapt the system’s reasoning logic in response to ongoing interaction. Similarly, Shaikh et al~\cite{shaikhRehearsalSimulatingConflict2024} use LLMs to simulate conflict resolution scenarios based on social conflict theory, helping users practice interpersonal skills in scripted and simulated environments. These systems signal a significant shift toward more proactive and theory-informed AI. However, they remain constrained by either single-user reflection, static expert knowledge, or fixed-role simulation. In contrast, domains like entrepreneurship coaching involve real-time, evolving interaction between asymmetrical roles in real-world problems, where novices and mentors bring different goals, knowledge, and constraints. Supporting this context requires systems that not only scaffold reflection but do so across multiple users, with adaptable, expert-editable reasoning structures embedded in live, high-uncertainty problem-solving contexts.

\section{Method}
To address our research questions, we adopted a \textit{Research through Design} (RtD) approach~\cite{zimmermanResearchDesignMethod2007b}, situated within an action research framework~\cite{hayesRelationshipActionResearch2011}. RtD is a design-oriented research approach widely used in HCI and CSCW to generate knowledge through iterative development and reflection on working systems in real-world contexts~\cite{zimmermanResearchDesignMethod2007b, stermanKaleidoscopeReflectiveDocumentation2023, gaverWhatShouldWe2012}. Rather than aiming to test hypotheses in controlled settings, RtD prioritizes situated experimentation and critical reflection to uncover new understandings about interaction, practice, and design. In particular, it emphasizes the productive tensions between problem-framing, system design, and empirical insight~\cite{zimmermanResearchDesignMethod2007b}---making it especially well-suited to studying complex, ill-defined domains like entrepreneurship coaching. 

We complemented RtD with~\textit{action research}, which focuses on iterative cycles of intervention and reflection, conducted in close collaboration with stakeholders in a real setting~\cite{hayesRelationshipActionResearch2011}. While RtD guided our approach to system design and generation of design knowledge, action research grounded our approach in a specific context---a university entrepreneurship incubator---where we collaborated closely with mentors and novices to iteratively adapt the system based on their evolving goals and needs. This combined approach allowed us to both intervene in practice and reflect systematically on the implications of our design decisions to broader theory and design principles. In the following sections, we describe the context of our research, the participants involved, and the specific methods used across the design and deployment phases.

\subsection{Context: University Incubator}
We chose to design and deploy a coaching system within a university incubator setting. During the past decade, university incubators have proliferated as institutions recognize their role in preparing the next generation of entrepreneurs~\cite{pittawayEntrepreneurshipEducationSystematic2007, pauwelsUnderstandingNewGeneration2016}. These incubators provide a unique context where novices tackle ill-defined problems lacking definitive solutions, well-defined problem statements, or specific problem-solving processes~\cite{jonassenEverydayProblemSolving2006, lynchDefiningIllDefinedDomains, huangIntelligentCoachingSystems2023}. Coaching is critical to developing entrepreneurial skills within these incubators~\cite{pittawayEntrepreneurshipEducationSystematic2007}. Novices in university incubators often pursue entrepreneurship as an extracurricular activity~\cite{huangIntelligentCoachingSystems2023}. However, the difficulty and burden of coaching in incubators are high due to the increasing number of novices, each working on different problems at varying paces and limited coaching resources~\cite{mianTechnologyBusinessIncubation2016, huangIntelligentCoachingSystems2023}. This setting poses significant challenges as novices work on diverse projects at different stages, making personalized support difficult. Designing a human-AI system to support coaching in university incubators can offer valuable insights for other ill-defined domains, including STEM inquiry learning environments~\cite{koedingerNewPotentialsDataDriven2013, linnInquiryLearningOpportunities2018}, research labs~\cite{zhangAgileResearchStudios2017}, knowledge work~\cite{samroseMeetingCoachIntelligentDashboard2021}, and medical education~\cite{hmelo-silverGoalsStrategiesProblembased2006, hmelo-silverFacilitatingCollaborativeKnowledge2008}.

\subsection{Participants}
We recruited four mentors and four novices from the university incubator for the first phase of our study to design and test low-fidelity prototypes. The four mentors included two women and two men, ages 24-44, who had 3-10 years of experience coaching entrepreneurs. The novices (ages 18-28) included three undergraduate students and one graduate student. Only one had prior entrepreneurial experience; the others were first-time founders.  

For the deployment, we recruited one mentor and 11 novice entrepreneurs (different from those in the design phase) from the same incubator during the summer 2024 term. Of all the novices, 4 were undergraduate students, 6 were master's students, and 1 was a Ph.D. student. Three had experience funding other ventures or working for startups before; the other eight were first-time founders. Furthermore, 7 of them had previously been coached by the mentor. The recruited novice population was sufficient for this study as it covered a wide range of experiences, stages, project areas, and previous experience with the mentor. This widespread was representative of the typical spread in this incubator and would allow us to understand how the system might support different novices thoroughly. The novices and the mentor used the coaching system for their real coaching meetings, each lasting 30 minutes to 1 hour. The IRB of the university approved this study. Before the study, all participants agreed to participate and allowed their data to be anonymously used for research purposes. They consented to audio, video, and screen recordings; audio recordings were transcribed for analysis. We conducted all sessions over Zoom video calls. Each participant was compensated with an Amazon gift card for their time. 

\subsection{Initial Design Principles}
Before designing our prototypes, we looked at the medical context for inspiration. Like entrepreneurship mentors, doctors need to gather extensive information on patient symptoms and medical history before they can accurately diagnose the causes of patient illnesses. But doctors often have a large number of patients to support and limited time with each of them~\cite{gholamzadehAppliedTechniquesPutting2021}. Instead, nurses will first meet with the patient to discuss their symptoms and review their medical history to ensure that the limited time between the doctors and the patient is spent effectively~\cite[e.g.,][]{niMANDYSmartPrimary2017, kimExploringNursesMultitasking2023}.

\begin{figure}
    \centering
    \includegraphics[width=0.9\linewidth]{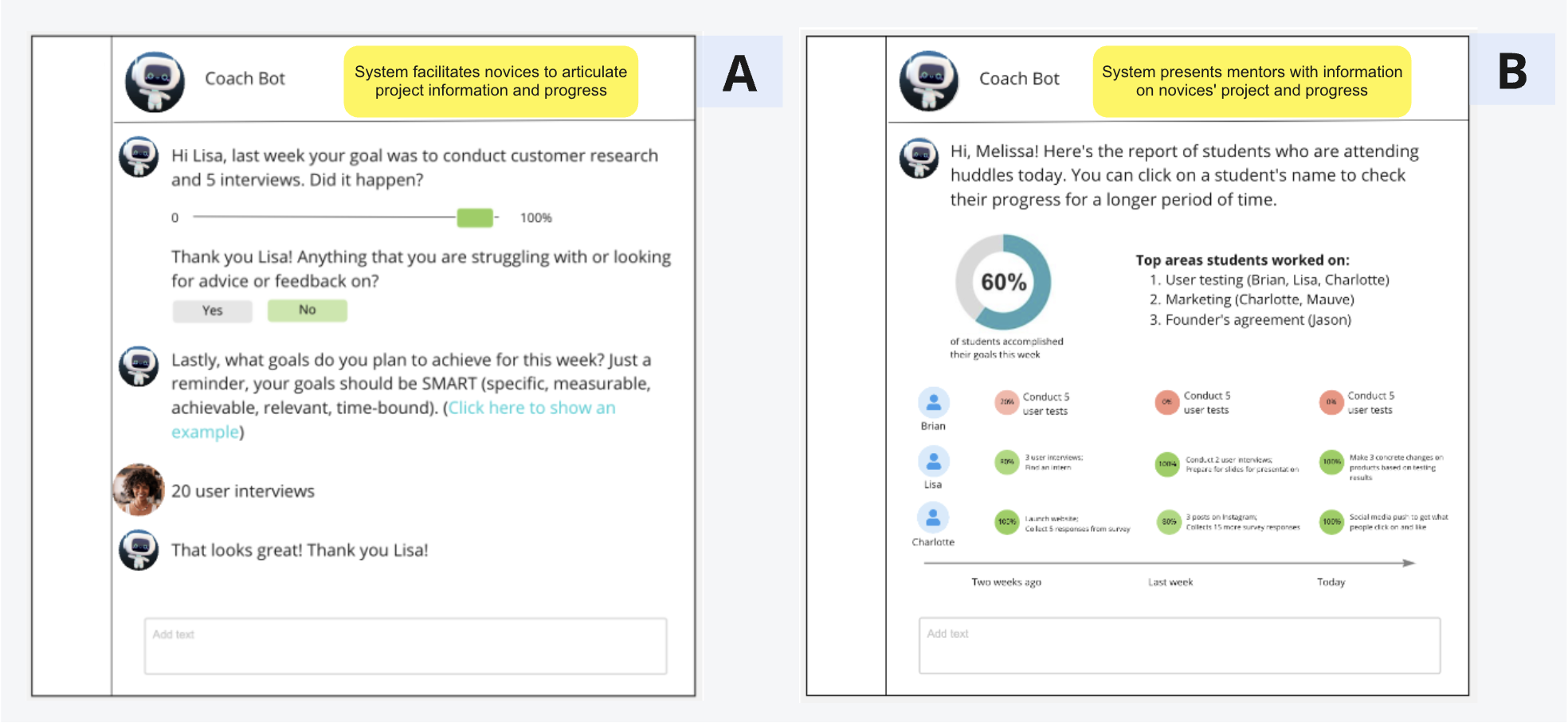}
    \caption{Examples of a prototype of a coaching system that we designed and tested with mentors and novices. This prototype system (A) asks novices questions about their projects and progress before meetings, and (B) presents mentors with summarized information on novices' projects and updates. }
    \label{fig:s3-prototypes}
\end{figure}

We hypothesized that if a coaching system can facilitate novices to provide information on their projects and progress before meeting with their mentors, it will help mentors and novices have more effective and productive meetings. Since the system has already performed the task of gathering information, mentors will have more time to spend diagnosing design risks and helping novices address those risks. Based on this hypothesis, we designed low-fidelity prototypes of a coaching system that facilitates novices to articulate project information and progress through a chatbot interface (Figure~\ref{fig:s3-prototypes}-A) before meetings and then presents this information to mentors through a dashboard (Figure~\ref{fig:s3-prototypes}-B) before coaching meetings. The initial questions that the chatbot helps novices to answer are based on common strategies that mentors use during coaching meetings ~\cite{huangIntelligentCoachingSystems2023}.

We iteratively tested and improved the prototypes with a total of four mentors and four novices to explore what information the AI should gather from novices, whether novices could provide this information, and how that information should be presented to mentors to maximize its utility during coaching meetings. In each test, we presented the prototype to the user and instructed them to imagine using it in a realistic scenario. We took notes on the user behavior and asked for their feedback on the desirability and usefulness of the prototype to facilitate a more effective and productive coaching meeting. We then improved the prototype based on user feedback before the next round of testing. The fidelity of the prototype increased over time from paper mock-ups with fabricated data to interactive mock-ups with real data.

User feedback highlighted the need for deeper and more personalized support for both novices and mentors (Table ~\ref{tab:initial-design-principles}). The mentors requested more control over the chatbot questions to better align with their coaching goals and wanted the system to guide the novices toward deeper reflection rather than simple status updates. The novices emphasized the importance of intrinsic motivation to use the chatbot and preferred questions tailored to their specific project needs. In addition, mentors sought rich contextual information from novice updates and reflections, but found too much raw information overwhelming.

\begin{table*}[t]
\small
\centering
\resizebox{\textwidth}{!}{%
\begin{tabular}{|p{3.8cm}|p{4.2cm}|p{4.5cm}|p{4.5cm}|}
\hline
\rowcolor[HTML]{EFEFEF} 
\textbf{(A) Initial design principles} &
\textbf{(B) Prototype features }&
\textbf{(C) User needs discovered} &
\textbf{(D) Design changes} \\ \hline

Novices articulate project information and updates before meetings &
Chatbot (Figure~\ref{fig:s3-prototypes}-A) prompts novices to reflect on project areas before meetings using questions derived from an existing coaching model~\cite{huangIntelligentCoachingSystems2023}. By mirroring mentor-like inquiry, the system can offload pre-meeting information gathering. &
\textbf{Mentors:} (1) Wanted control over system-posed questions; (2) Preferred deep reflections, not just reports. \newline
\textbf{Novices:} (3) Needed tangible value to stay engaged; (4) Desired personalized, context-aware support. &
(1) Enable mentor to modify system questions. \newline
(2) System diagnoses project risks and prompts reflection on risks to add value and scaffold deep reflections. \newline
(3) LLM tailors questions to novices' project context.\\ \hline

Mentors review novices’ project information and updates before meetings &
Dashboard (Figure~\ref{fig:s3-prototypes}-B) shows a structured summary of project updates and context to reduce time spent on status updates and increase time for diagnosis and strategy. &
\textbf{Mentors:} Desired rich, contextualized insights without cognitive overload. &
(1) LLM summarizes novice input to reduce overload. \newline
(2) Novices prioritize needs to direct mentor focus. \\ \hline

\end{tabular}
}
\caption{Through iterative prototyping and testing our initial design principles (A), we discovered user needs (C) and proposed concrete design changes (D) to address those needs.}
\label{tab:initial-design-principles}
\end{table*}

\subsection{A Proactive Human-AI Coaching System}

Based on the initial design principles and discovered user needs (Table~\ref{tab:initial-design-principles}, Column A\&C), we developed a human–AI coaching system that supports entrepreneurship coaching through proactive, expert-governed, and relational scaffolding. The system integrates a LLM with a cognitive coaching model~\cite{carlsonDesignRisksFramework2020, huangIntelligentCoachingSystems2023} to provide differentiated and proactive support to both novice entrepreneurs and mentors. Under the hood, the system is embedded with a coaching model that has two key knowledge structures: (1) A project model that outlines relevant project areas that novices should articulate on, such as the problem, current focus, and plans (Appendix~\ref{tab:s3-projquestions}); and (2) a risk model that outlines common design risks such as vague understanding of intended users' needs and not having evidence on how proposed solutions will address those needs (Appendix~\ref{tab:s3-riskmodel}). 

\begin{figure}
    \centering
    \includegraphics[width=\textwidth]{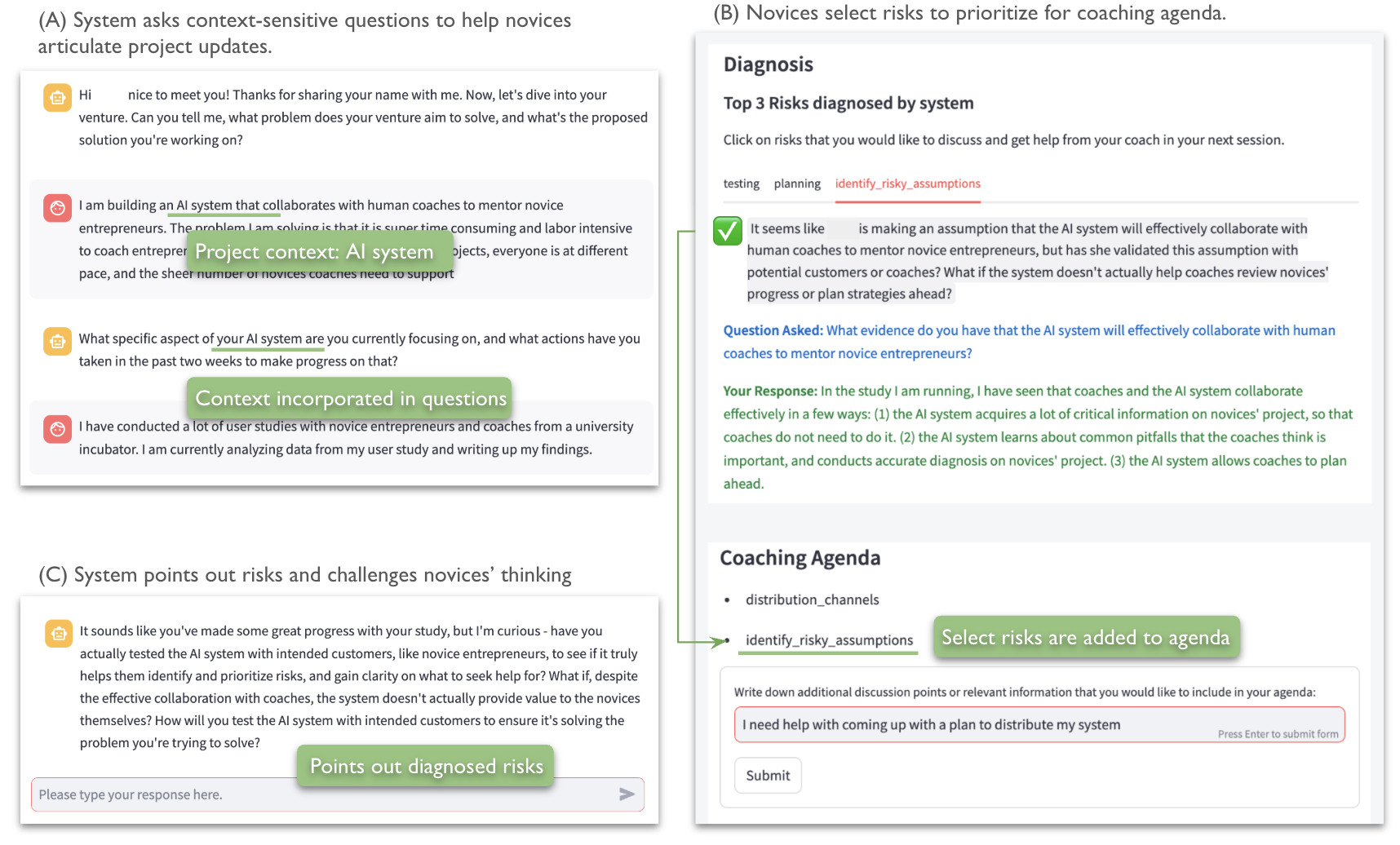}
    \caption{When supporting novices, the coaching system operates through two interfaces: a chatbot (A\&C) and a coaching dashboard (B). Through these interfaces, the system offers adaptive and proactive support for novices: it guides novices to articulate, diagnose, and prioritize risks. }
    \label{fig:novice-interfaces}
\end{figure}

\begin{figure}
    \centering
    \includegraphics[width=\textwidth]{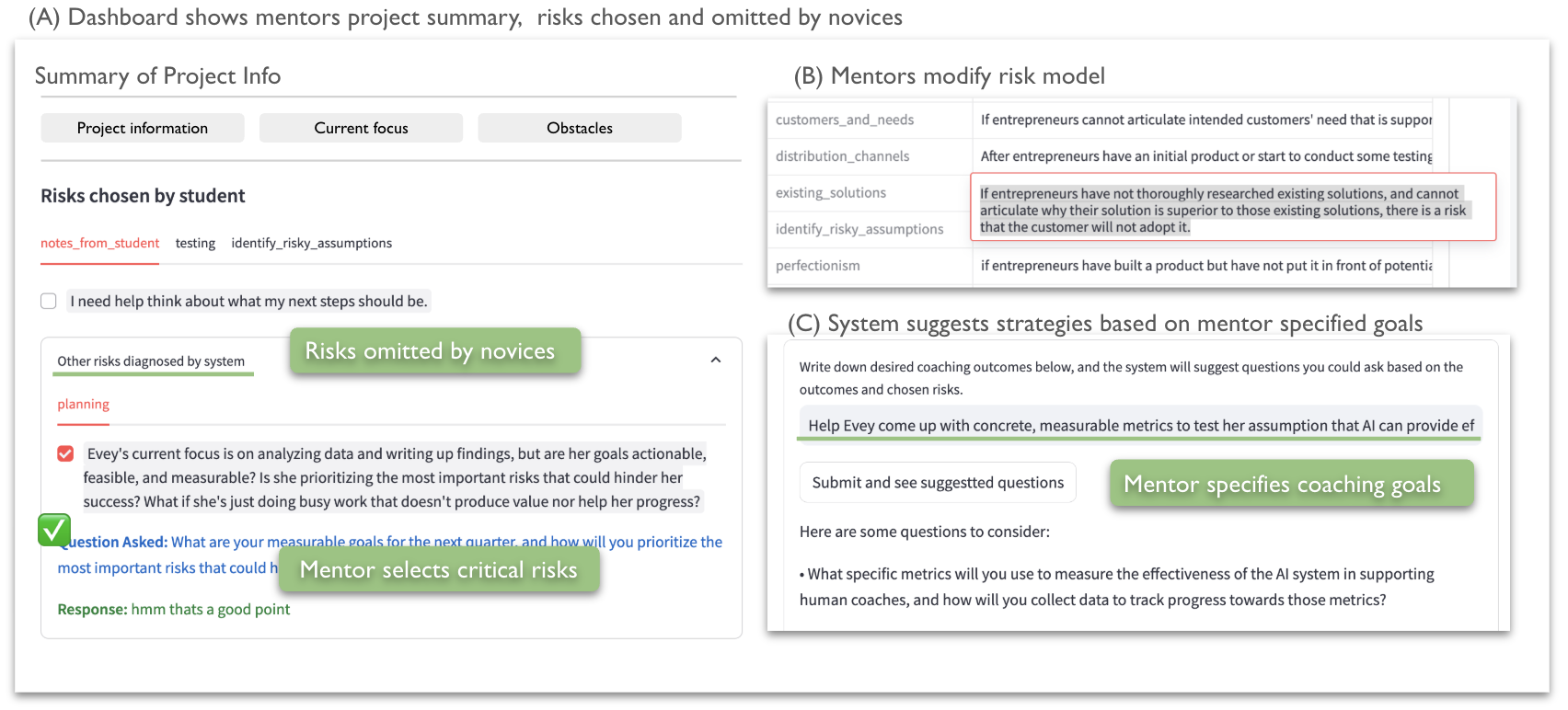}
    \caption{When supporting mentors, the coaching system operates through two interfaces: a coaching dashboard (A\&C) and an authoring interface (B). The dashboard shows mentors a summary of novices' project updates and risks both chosen and omitted by novices. It also suggests coaching strategies based on risks and coaching goals specified by mentors. Using the authoring interface, the mentor can inspect or modify existing risks and add additional risks. }
    \label{fig:mentor-interfaces}
\end{figure}

\subsubsection{System features and interfaces}
The implemented system consists of several integrated components designed to scaffold novice reflection, support mentor preparation, and adapt dynamically to evolving project contexts. Before each coaching meeting, the system engages novices through a chatbot interface that guides them to articulate key areas of their projects and reflect on system-diagnosed risks (Fig~\ref{fig:novice-interfaces}-A). Using a LLM and based on a cognitive coaching model~\cite{huangIntelligentCoachingSystems2023, carlsonDesignRisksFramework2020}, the system generates questions aligned with key project areas, such as intended problems and customers' needs, outlined in the model. These questions also adapt in real time based on the project context provided by novices. For example, rather than using a generic prompt such as "what is your current focus?", the system uses a novice's project context, an AI coaching system, to personalize the question into {~\em``What specific aspect of this AI system are you currently focusing on?''}. Once the system gathers responses across all project areas, it applies the risk model to identify potential design risks. The system continues the dialogue by surfacing each diagnosed risk and asking reflective questions tailored to novice project contexts, encouraging deeper thinking about project uncertainties and assumptions (Fig~\ref{fig:novice-interfaces}-C).

Following the chatbot conversation, the system generates a dashboard for novices that includes a summary of their project information, detailed reports of diagnosed risks, and a space for adding additional notes and agendas (Fig~\ref{fig:novice-interfaces}-B). This dashboard explains each risk and the associated reflective questions and responses, and allows novices to prioritize specific risks for the upcoming coaching meeting. By surfacing other risks from the model and giving novices space to note concerns, the system encourages self-assessment and grants novices greater agency in shaping meeting agendas.

Mentors access a parallel dashboard that summarizes the status and reflections of the novice project (Fig~\ref{fig:mentor-interfaces}-A). This includes novice-prioritized risks along with those left unselected, offering mentors insight into what novices value and what might require additional attention. The dashboard shows the system's rationale for each diagnosis and includes full transcripts of chatbot interactions, allowing mentors to independently verify the assessments. In addition, mentors can indicate which risks they want to focus on and specify desired outcomes of the meeting. Based on these inputs, the system uses an LLM to generate coaching strategies that are both risk- and context-specific, aligning with the mentor’s goals and the novice’s project stage (Fig~\ref{fig:mentor-interfaces}-C).

To support continual refinement, the system allows mentors to inspect and modify the underlying risk framework (Fig~\ref{fig:mentor-interfaces}-B). While the initial model draws from established frameworks~\cite{carlsonDesignRisksFramework2020, huangIntelligentCoachingSystems2023}, mentors can revise existing risks or add new ones based on emerging patterns observed in practice. This ensures that the system remains adaptable and responsive to diverse coaching needs over time.

\subsubsection{Implementation Details}
The coaching system is developed using (1) Python for backend logic, (2) Streamlit as the front-end interface, (3) LangChain with the LLaMA 3–70B language model for language understanding and generation, and (4) Firebase as a real-time cloud database for storing evolving user context and expert knowledge.  

At its core, the system leverages a structured cognitive coaching model~\cite{carlsonDesignRisksFramework2020, huangIntelligentCoachingSystems2023} and evolving user context to guide the behavior of the LLM. This model encodes expert mentor knowledge into two hierarchical structures: (1) a project model that outlines key areas (e.g., user needs, problem framing, solution assumptions, experimentation plans) that novices should articulate on as contexts; and (2) a risk model that maps common novice pitfalls (e.g., designing for imagined users, skipping validation, or working on low-priority tasks). These models are stored as JSON files in Firebase and can be updated by mentors through a low-barrier front-end editor. 

The system incorporates this expert model into a sequence of prompts that guide the LLM through different reasoning tasks: (1) project information extraction, (2) risk diagnosis, (3) reflection question generation, (4) strategy suggestion for mentors, and (5) agenda synthesis. Each prompt template incorporates three layers of information: 
\begin{enumerate}
    \item \textbf{Domain knowledge} from the coaching model: definitions of project areas, common risks that novice founders encounter, and concrete examples of those areas and risks.
    \item \textbf{Project-specific context} is extracted from the novice's inputs, such as user problems, updates on goals, and next steps. These are structured into key-value pairs (e.g., \textit{Goals: "Finish onboarding one hundred users"}) and passed into prompts via LangChain’s input variable bindings.
    \item \textbf{Task-specific prompt framing}: tailored system instructions for different reasoning tasks (e.g., ~\textit{"Given the provided project context and definitions of common risks, identify which of the following risks may apply. Justify your reasoning using the definitions and examples provided"}).
\end{enumerate}

\begin{figure*}[!t]
\centering
\includegraphics[width=\textwidth]{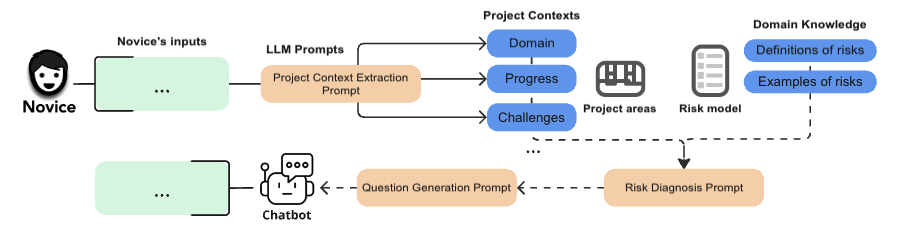}
\caption{The system incorporates project contexts extracted from novices' inputs and domain knowledge on risk definitions and examples to diagnose potential risks and generate targeted, reflective questions.}
\label{fig:chained_prompts}
\end{figure*}

Each step in the prompt chain passes structured intermediate outputs (e.g., extracted project updates, identified risks) to the next stage (see Figure~\ref{fig:chained_prompts}). For instance, after the LLM diagnoses a set of project risks, the next prompt asks it to generate targeted follow-up questions that challenge the novice’s assumptions related to those risks. More details on each prompt are included in the appendix.

Mentors can inspect and modify the coaching model via a structured table-based interface (see Figure~\ref{fig:mentor-interfaces}-B). Changes such as adding a new risk category or revising an example statement are saved to Firebase and immediately reflected in the next system interaction. This makes the LLM behavior transparent and governable, transforming it into an adaptive, inspectable, and co-designed reasoning partner.

\subsection{Exploratory Field Deployment}
To understand how this human-AI coaching system supports mentors and novices to have more effective and efficient meetings, we conducted an exploratory field deployment of the coaching system in the same entrepreneurship incubator community. An exploratory field deployment is suitable to help us understand the impact of the coaching system within the intended context of use~\cite{siekFieldDeploymentsKnowing2014}. This method is also appropriate to evaluate the system that has been developed using the RtD methodology~\cite{zimmermanResearchDesignHCI2014, zimmermanResearchDesignMethod2007b} that has been guiding this research. Specifically, we conducted a semi-controlled deployment in which 11 novice entrepreneurs and one mentor used the system to facilitate their real coaching meetings. 

\subsubsection{Procedure and Data Collection}
We collected much data, including interviews, surveys, and video recordings. This breadth of data allowed us to learn, understand, and triangulate how our human-AI coaching system that uses a cognitive model and an LLM affects novices, mentors, and coaching meetings. 


For novices, data collection began with a pre-system interview to establish a baseline understanding of their project plans, perceived risks, and coaching goals. During their interaction with the system, we recorded screen-sharing sessions and asked them to think aloud. Post-interaction interviews explored how the system influenced their diagnostic reasoning. Novices also completed Likert-scale surveys evaluating the relevance of the system-diagnosed risks and the perceived value of the experience. After the coaching session, a follow-up interview asked them to reflect on the session’s usefulness and how it differed from prior meetings without system support.

For the mentor, we recorded think-aloud sessions as he reviewed each novice’s dashboard before their meeting. We interviewed him afterward about the accuracy of system-diagnosed risks, the clarity of his meeting plans, and his overall experience using the system. A second interview followed each coaching session to assess how the system shaped the interaction and whether it improved upon prior meetings. Finally, each coaching session---lasting 30 minutes to 1 hour---was conducted 1-on-1 between the mentor and the novice without researcher presence. These meetings were video recorded and reviewed later as part of our analysis.

\subsubsection{Measures and Analysis}
We conducted a thematic analysis~\cite{Braun2006} to examine the interview data, focusing on how the coaching system supported novices and mentors. To understand whether and how the system supports novices' meta-cognitive processes, we identified instances where novices explained how the system aided their articulation, diagnosis, and planning. To understand how the system supports mentors in coaching novices, we noted how mentors shared their experiences using the system in diagnosis and planning for the meetings. We also looked for information on how the system affected coaching meetings compared to previous experiences. Throughout our analysis, we captured the challenges faced by both novices and mentors, refining our themes by analyzing supporting and contradictory data until each theme was distinct. The lead author coded the transcripts and generated initial themes, which were collectively refined and finalized by all authors.

We analyzed both the interview data and the survey results to assess the precision of the system's diagnoses. The novices rated risks on a 5-point Likert scale for relevance, criticality, and actionability. The mentors also evaluated these risks during interviews. Additionally, we used objective measures to evaluate the system's impact. We compared novices' plans and risk identifications before and after using the system and similarly examined the mentors' diagnoses pre- and post-meeting. We wanted to see whether the risks identified by the system remained relevant after discussions and whether the mentor identified additional risks, indicating possible insufficiency in the system diagnoses.

\section{Findings}

In this section, we present the findings of the deployment study, organized around five design principles that emerged through our thematic analysis. These principles reflect how the system supported the interactions of novices and mentors with each other and AI. As in previous work, we found that several design choices simultaneously enabled meaningful support while also introducing new forms of friction or dependency. Table \ref{tab:final-design-principles} summarizes these emerging design principles, along with the specific features of the system that instantiated them, and highlights the key successes and challenges observed during deployment.

\begin{table*}[t]
\small
\centering
\resizebox{\textwidth}{!}{%
\begin{tabular}{|p{3.5cm}|p{4.5cm}|p{4.5cm}|p{4.5cm}|}
\hline
\rowcolor{lightgray} 
\textbf{(A) Final design principles} & \textbf{(B) System features}& \textbf{(C) Findings: Successes}& \textbf{(D) Remaining challenges}\\ \hline

\textbf{1. Proactively challenge novices’ thinking on risks}& 
The system proactively surfaces risks and poses pointed, diagnostic follow-up questions to challenge novices' thinking. & 
\textbf{Novices:} Using the system helped uncover blind spots, confront neglected risks, and prioritize critical risks. & 
\textbf{Novices:} Some distrusted the system when challenged; others expected answers rather than proactive questions due to existing experience with LLM chatbots. \\ \hline

\textbf{2. Layer dual context to adapt support}& 
The system adapts support using both novices' inputs and mentor-specified goals as contexts. It aligns AI scaffolding with shared human priorities and evolving needs.& 
\textbf{Novices:} Diagnoses and questions tailored to their project contexts made risks feel more “real” and the interaction more engaging. \newline \textbf{Mentor:} Strategy suggestions that have taken into account his coaching goals were more specific and actionable.& 
\textbf{Novices:} Some provided insufficient context, leading to inaccurate diagnoses and additional investigation from the mentor.\\ \hline

\textbf{3. Empower mentors to govern domain-grounded AI logic} & 
The system’s core reasoning is based on a cognitive coaching model that encodes how expert mentors reason about risk, adapt strategies, and diagnose root causes across diverse, evolving ventures~\cite{huangIntelligentCoachingSystems2023, carlsonDesignRisksFramework2020}. Mentors can inspect and modify this model through a low-barrier authoring interface---adding new risks, revising diagnostic logic, and tailoring strategies---all without coding.& 
\textbf{Mentor}: modified the model easily; 
\newline \textbf{Both:} found the system diagnoses relevant and aligned with coaching goals. & 
Building the initial model was time-consuming; the current version supports only one mentor and limited stages of projects.\\ \hline

\textbf{4. Surface root-causes---both emotional and cognitive---to scaffold mentor insight}& 
The dashboard shows novices’ reflective responses, risk omissions, and agenda selections to scaffold mentor insight into root causes of risk. These patterns surface not just deeper beliefs but emotional blockers like fear of failure, perfectionism, or overconfidence---barriers often invisible in traditional scaffolds. & 
\textbf{Mentor:} Recognized root causes, including emotional blockers, accurately when reviewing the coaching dashboard and adjusted his focus and strategies for the meeting. 
\newline \textbf{Novice:} Felt they could be more honest with the system. & 
\textbf{Mentor:} Verified dashboard information to avoid the risk of hallucination.\\ \hline

\textbf{5. Orchestrate human-human collaboration through asynchronous, role-sensitive preparation}& 
The system prepares both mentor and novice ahead of meetings by prompting novices to articulate risks and generate an agenda and providing mentors with a synthesized dashboard and strategy suggestions. Using the system lays the groundwork for collaboration before it begins. & 
\textbf{Novices:} gained clarity on needs and strategized for discussion. \newline \textbf{Mentor:} felt more “prepared and intentional.” \newline  \textbf{Both: } Considered meetings to be more efficient and in-depth and engaged in productive disagreement. & 
\textbf{Novices: } Some initially misunderstood the system’s collaborative role---expecting it to behave like a reactive AI assistant rather than a preparatory tool for human-human interaction. \\ \hline

\end{tabular}%
}
\caption{Summary of design principles, system features, and deployment findings to inform future human-AI coaching systems for ill-defined domains}
\label{tab:final-design-principles}
\end{table*}

\subsection{Proactively Challenge Novices to Support Critical Thinking on Risks}
In ill-defined domains like entrepreneurship, novices often lack the expertise and meta-cognitive skills to identify what they do not know~\cite{reeslewisAssessingIterativePlanning2019, carlsonDesignRisksFramework2020}. The system was designed not to simply follow a novice’s framing but to gently challenge assumptions, confront blind spots, and ask necessary but uncomfortable questions. Drawing from a cognitive coaching model and project-specific context, it proactively exposed potential risks, even those not recognized by the novice, and asked pointed diagnostic follow-up questions~\cite{sarkarAIShouldChallenge2024}.

In several cases, the system helped to uncover blind spots by bringing attention to risks that novices had not previously considered (P1, 7, 8, 10). After P8 described her venture connecting artists to local fairs, the system diagnosed a risk related to distribution channels and asked {\em``How do you plan to get your products into the hands of artists? And what evidence do you have that this strategy might work?''}. P8 later admitted that this was something she had not considered before.  Similarly, P1 reflected on how the system highlighted risks that he was not previously aware of, offering a new perspective on his project. 

Even when aware of critical risks, novices often gravitated toward tasks that felt easier and more manageable (P1, 2, 5). By proactively pointing out risks and challenging novices, the system helped them reorient their priorities (P1, 2, 3). When the system diagnosed that P1 lacked a distribution plan, he reflected: {~\em``I felt called out---but in a good way.''} He acknowledged that while he was aware of this risk before, he prioritized easier tasks like app development, which made him feel productive. The diagnosis served as an {~\em``immediate reminder''} of what was more important.  Similarly, P2 said the system posted {~\em``pivotal questions''} (P2) that helped her confront risks she had been avoiding, ultimately leading to a deeper understanding of her situation. The proactive nudging from the system led novices to shift their focus away from comfortable tasks (like app development or metrics creation) and toward more critical but cognitively demanding challenges. 


This proactive behavior extended beyond traditional scaffolding tools such as templates. P8 described that while she appreciated how templates helped structure ideas, the system scaffolded her understanding by showing critical risks that might not have been immediately apparent: {\em``A lean canvas might not show holes in my arguments. It might look like a complete picture, whereas the system is going a layer deeper''}. The system was able to go {~\em``a layer deeper''} (P8) in identifying risks and guiding her through the diagnostic process in a way that static templates such as business model canvases did not. 

However, not all users interpreted the proactive approach of the system positively. Some novices (P6, P7) distrusted the system’s judgment unless it aligned with their thinking. P7 remarked, {~\em``I don’t know if I would trust its opinions… because it’s a chatbot.''} Others were caught off guard by the system’s questioning style, having expected it to behave more like ChatGPT by offering direct advice (P7, P8, P10). This expectation mismatch sometimes caused initial confusion about the system’s purpose as a reflective rather than an instructive tool.

\subsection{Layer Dual Context to Adapt Support for Both Novices and Mentors}
The system adapts its support based on (1) the novice's articulated context (e.g., project goals, assumptions, and reflections) and (2) the mentor's specified coaching goals and concerns. This dual-context approach shapes both the risk diagnosis for novices and the coaching strategy suggestions for mentors. Personalized support in ill-defined domains requires understanding evolving needs from both sides of an interaction~\cite{huangIntelligentCoachingSystems2023, gargOrchestrationScriptsSystem2023, reeslewisBuildingSupportTools2015}. By layering both novice and mentor perspectives, the system enables context-aware AI support that aligns with shared priorities and prevents one-size-fits-all advice.  

This dual-context approach introduced a level of adaptability that is absent in traditional scaffolds.  For example, after learning that the novice is working on a fantasy league game for pickle-ball players, it adapted a general risk---such as {\em``untested assumptions may hinder growth''}---into a risk specific to the context, such as whether the novice had tested his assumption that the concept of such a game {\em``would resonate with pickle-ball players''} (P3). Novices found the domain-specific and context-aware questions more engaging and specific, prompting deeper reflection on their ventures (P2, 3, 7, 8, 9). P9 observed that she might write {~\em``nobody buys my product''} in response to a general question like {~\em``what are the most critical risks''} on a canvas. But the system personalized its diagnosis question with P9's project contexts, asking how she might test customer willingness to pay for a nostalgia-themed beauty product (the core selling point of her venture). As a result, it pushed her to think {~\em``more deeply.''} 

Furthermore, the system's ability to refer to previous context during interactions added another layer of adaptivity and personalization, making the experience more {\em ``engaging''} for novices (P1, P7). P7 explained that while the questions were similar in substance, the system’s one-at-a-time approach and use of her previous responses made the process feel {\em ``a lot less intimidating''} and more interactive. As a result, novices could approach their tasks with greater focus and confidence, ultimately enhancing their learning and reflection. 

This personalization also extended beyond novices to support mentor needs. Initially, the system generated coaching strategy suggestions based solely on novice project updates and diagnosed risks. The mentor found these suggestions {\em``too broad''}, generic, and not immediately actionable. After the system began to incorporate the mentor's coaching goals - collected through a lightweight authoring interface (Fig~\ref{fig:mentor-interfaces}-B) - the suggestions became much more relevant and useful. The mentor noted that the revised suggestion often surfaced issues he had not considered, stating: {~\em``I probably wouldn’t have thought to ask that question.''} He emphasized that personalized, context-rich suggestions allowed him to enter meetings better prepared: {~\em``What would augment my job is giving me exact questions or topics to discuss, rooted in the context of that person’s story and what they shared with the system.''}

Despite its personalization capabilities, the precision and usefulness of the system were limited by the quality and completeness of the context it received. Novices who interacted with the system in brief, generic terms---often out of habit from using tools like ChatGPT---received shallower diagnoses. P7 admitted to providing minimal input because she was used to writing short prompts for other LLMs. P4 and P6 encountered similar problems: the system flagged risks they had already addressed, not because it misunderstood the problem but because they had not yet disclosed the relevant context. P4 reflected, {~\em``It’s less a problem of the system misdiagnosing and more about it not having enough information.''} P6 said, {~\em``It doesn’t know the full context of what my skills and past experience are… so I’m okay with the fact that it’s not perfect.''} These examples illustrate that the system's ability to offer personalized support is fundamentally limited by what users choose to share---and are asked to share. Designing interfaces that actively encourage richer, more structured self-disclosure may be key to unlocking more adaptive and meaningful support.

When the information on the dashboard was lacking context or clarity, the mentor also had to ask additional questions during the meetings to fill in the gaps. For example, when reviewing the P6 dashboard, the mentor struggled to determine the accuracy of the system-diagnosed risks because it lacked important contextual details, information that novices had not articulated during their interactions with the system. Key details, such as who the users and stakeholders were and what relationships P6 currently had with them, were missing. Similarly, he reported feeling {\em``ill-informed''} when reviewing P2’s dashboard due to a lack of clarity, which hindered his understanding. As a result, the mentor had to ask multiple questions during the meeting to gather the necessary context and clarify the information presented.

\subsection{Empower Mentors to Govern Domain-grounded AI Logic}
The system’s core reasoning is built from a cognitive coaching model constructed through in-depth observation and analysis of real mentoring sessions in a university incubator~\cite{huangIntelligentCoachingSystems2023, carlsonDesignRisksFramework2020}. This model encodes how expert mentors reason about risk, adapt strategies, and diagnose root causes in diverse, evolving ventures. Mentors can inspect and modify this model through a low-barrier authoring interface: adding new risks, rewriting diagnostic logic, and tailoring strategies, all without coding (Fig~\ref{fig:mentor-interfaces}-B). Rather than treating AI behavior as static or opaque, the system invites mentors to critique, evolve, and coauthor its logic. This supports trust, adaptability, and alignment in ill-defined domains such as entrepreneurship, where mentorship is relational, situated, and contingent.

As a result, the system successfully identified risks that both novices and mentors found relevant and critical. Many novices responded to the system’s diagnosis with comments like {\em``This is exactly what I am struggling with''} (P3, 8), while the mentor often confirmed by saying, {\em``this is spot on''} or {\em``this is the biggest risk''}. In their survey responses, novices rated the relevance and criticality of system-diagnosed risks highly, indicating that the system effectively identified critical areas that needed attention. 

When the system overlooked important risks that were not represented in the model, the mentor was able to modify the risk framework directly using the authoring interface. These updates did not require technical skills and could be completed in minutes. For example, during the session with P8, the mentor diagnosed a risk with teamwork: the co-founders were pursuing separate ideas. But the system did not identify this risk because there was no articulated risk around teamwork in the risk model. The mentor immediately added this risk using the interface after the meeting. A similar addition occurred for P7, who was struggling with leadership responsibilities following the departure of several teammates. These quick adjustments helped ensure that the system could more accurately reflect the realities of each novice's venture in future sessions.

In addition to adding new risks, the mentor refined the phrasing of existing ones to improve diagnostic specificity. Initially, a risk was framed around whether the novice had a plan for regular customer testing. Over time, the mentor observed that assessing the depth and quality of the testing strategies, such as what was being tested, how, and with whom, was more critical than simply whether testing occurred. He immediately revised how the risk of testing was articulated in the model to reflect this change. These edits, made through an editable table (Figure~\ref{fig:mentor-interfaces}-B), allowed the mentor to align the logic of the system with his evolving coaching goals.

The participants also contrasted the behavior of the system with general-purpose AI tools such as ChatGPT or Claude. Several novices noted that these tools provided overly broad or generic advice when asked about entrepreneurial risks (P4, P6, P9, P10). For example, P9 described the responses of ChatGPT as {~\em``general information about starting a business,''} while the system asked specific questions about her assumptions, such as the willingness of the customer to pay for a nostalgia-based beauty product. P6 explained that what made the system more valuable was the integration of mentor expertise and context, something generic chatbots lack. These comparisons highlight the importance of making AI systems grounded in expert models and controllable by the mentor rather than static or purely generative.

After using the system, the mentor pointed out  a key limitation: the current risk diagnosis logic did not sufficiently account for variation in project types and stages. For instance, he noted that a strong emphasis on fundraising might represent a critical risk for a software startup---where upfront capital needs are often low---but not for hardware ventures, which typically require early investment to build physical prototypes. This example underscores the challenge of generalizing risk models across diverse entrepreneurial contexts. Moreover, the cognitive model embedded in the system required significant time and effort to construct~\cite{huangIntelligentCoachingSystems2023,carlsonDesignRisksFramework2020}. Prior work has shown that building such a model involves extensive ethnographic observation and analysis; for example, Huang et al~\cite{huangIntelligentCoachingSystems2023} observed over 20 hours of recorded coaching sessions to extract mentoring strategies and infer the reasoning processes behind them. Although this effort is substantially lower than that required to build ITSs~\cite{vanlehnAndesPhysicsTutoring2005,butzWebBasedIntelligentTutoring2004}, it still raises concerns about the scalability and transferability of this approach to new domains. 

\subsection{Surface Root-causes---Both Emotional and Cognitive---to Scaffold Mentor Insight}
The system invites novices to reflect on diagnosed risks, selects critical risks for meeting agendas, and shows mentors both the risks novices prioritized and those they omitted. This scaffolding allows mentors to interpret patterns of concern, avoidance, or inconsistency, surfacing potential root causes such as perfectionism, fear of failure, or misaligned mental models (e.g., building based on personal preferences instead of customer needs). This approach respects user agency and avoids intrusive inference while still giving mentors access to deeper signals. It mirrors how mentors read between the lines and piece together root causes based on what is said---and what isn’t. Rather than automate diagnosis, the system supports sense-making and empathy in high-stakes relational interactions.

The system successfully surfaced both cognitive and emotional root causes by creating a space where novices felt more comfortable disclosing their struggles---insights that mentors often could not access through conversation alone. Several novices reported feeling more open and honest with the system than they typically were with mentors, due to a perceived lack of judgment. For example, P1 admitted he often avoided showing vulnerability in front of mentors, yet disclosed key challenges when interacting with the system. Similarly, the mentor noted being surprised by P3’s openness in disclosing his challenges when reading the dashboard reflections, which contrasted with P3's confident demeanor in past meetings. These moments provided the mentor with access to internal barriers---such as perfectionism or fear of appearing incompetent---that could otherwise remain invisible.
 
The dashboard allowed the mentor to identify {~\em``what is causing  [novices] the most amount of stress''} by interpreting patterns of risks that novices selected for their coaching agendas. In the case of P1, the mentor observed that several of the risks appeared interconnected and suggested they stemmed from a fundamental lack of understanding of the product’s value:{\em``There may be a lack of understanding of what it really is, or it may not be compelling enough... These hit on different facets of his product and company.''} By connecting these signals, mentors were able to uncover deeper, underlying causes that were not explicitly articulated by the novices.

Novices' responses to the diagnosed risks allowed the mentors to identify possible root causes, such as underlying assumptions. For example, P2 responded to a risk about unvalidated assumptions by saying,{\em``What I’m developing is more to create what I think is worth developing than simply what can please them''}, revealing a deeper mindset: prioritizing her preferences over customer needs. The mentor recognized this immediately: {~\em“You just told me why this thing isn’t working. We’ve got to discuss that.”} These moments show how novices' reflections---when prompted by the system---gave mentors access to novices' underlying beliefs and blockers that shaped behavior.

In addition to what the novices said, what they chose to omit from their coaching agendas also revealed emotional cues. In domains like entrepreneurship---where personal identity, uncertainty, and self-worth are deeply entangled---emotional risks are often root causes of stagnation~\cite{huangIntelligentCoachingSystems2023} yet rarely surfaced through conventional checklists or templates. Externalizing tacit emotional cues can help mentors intervene with greater relational sensitivity and precision.

The mentor found the omitted risks on the dashboard {~\em``very telling''} of potential emotional blockers and adjusted strategies accordingly. When P1 omitted a risk related to testing assumptions, the mentor suspected fear of exposing the idea to criticism, which was later confirmed in the session as a fear rooted in past experiences of idea theft. In another case, P6’s omission of market adoption risks (which the mentor considered to be critical) led the mentor to hypothesize overconfidence: {~\em``This one is telling me he feels very confident people will adopt it.''} These observations enabled the mentor to interpret not just the novice's strategy but also their mindset and approach to the session with greater care. For P1, the mentor anticipated defensiveness and carefully framed the conversation to avoid triggering ego or fear:{~\em``I need to be careful how I navigate that... making sure I don’t trigger the ego or make them feel like they’re being attacked.''} This sensitive coaching strategy ultimately led to a breakthrough, as P1 acknowledged his fear and reconsidered how he should share his ideas. 

However, due to concerns about the system’s potential for {\em ``hallucination''}, the mentor often began sessions by verifying the information displayed on the dashboard. The mentor described how, similar to experiences with tools like ChatGPT or Claude, he did not fully trust the AI output to be completely accurate:{\em``I find myself wanting to verify that information. And I think that’s an interesting dynamic because that’s my experience when using Claude or ChatGPT... I don’t trust it to always be 100 percent. So I spend a good chunk of the early conversation just kind of asking questions to verify that information.''} This verification step was critical for the mentor to build confidence in the AI’s diagnosis and ensure the subsequent conversation was grounded in accurate, reliable insights.

\subsection{Orchestrate Human-Human Collaboration through Asynchronous, Role-sensitive Preparation}

Before a coaching meeting, the system supports each party asynchronously: novices receive prompts and diagnostic questions to articulate, reflect, and clarify needs, while mentors receive a dashboard summarizing novices' project updates, risks, and possible strategies tailored to mentors' coaching goals. Both parties then enter the meeting with a shared focus. This is especially important in domains like entrepreneurship, where mentoring is episodic, roles are asymmetric, and coaching depends on making the most of limited shared time~\cite{huangIntelligentCoachingSystems2023, reeslewisBuildingSupportTools2015}. An effective coaching meeting requires more than shared documents but \textit{shared readiness}~\cite{reeslewisBuildingSupportTools2015, huangIntelligentCoachingSystems2023, gargOrchestrationScriptsSystem2023}. By scaffolding both mentors' and novices' reflection and planning processes before joint interaction, the system increases intentionality, reduces cognitive load during meetings, and accelerates productive dialogue. It turns AI into a mediator, not just a tool.

Before meetings, novices articulated project contexts and reflected on the risks diagnosed with the system. This scaffolded preparation helped novices enter meetings with clearer priorities and talking points. For instance, P3 emphasized that preparation was key to receiving useful advice: {~\em``If I’m not able to convey my problem properly, their advice wouldn’t be helpful anyway.''} P8 noted that answering the system’s prompts helped her {~\em``hone in on the context,''} resulting in a more efficient and focused conversation. P2, P3, P7, and P8 reported that the system enabled them to clarify what to talk about and strategize how to present those concerns.

The system generated a structured coaching agenda based on selected and omitted risks. This helped novices frame their concerns with precision. P2 said the structured agenda made it easier to communicate with the mentor: {~\em``I can be more efficient and precise regarding what I want to talk about.''} P4, who had previously entered meetings with {~\em``vague concerns''}, found that being able to select risks made his needs more concrete: {~\em``Having something I can select that says all of that helps to know that’s an option I can talk about.''} P7 also described how the agenda made her progress more {~\em``digestible''} to the mentor, and P8 observed that it enabled both parties to {~\em``focus on the most relevant challenges''} and engage in a {~\em``concentrated effort.''}

The mentors reviewed a dashboard summarizing the updates of each novice, the selected risks, and the system-generated strategy suggestions before meetings. This transformed the mentor’s role from real-time problem-solver to intentional strategist. He described how, in traditional coaching, he often felt forced to analyze, synthesize, and strategize in real time, stating, {~\em``There is just so much going on in your brain when you're talking to someone... The day you get the information is not the day you can probably decide.''} With system support, however, he was able to outline targeted strategies in advance. For instance, after reviewing P1’s dashboard, the mentor planned to examine testing tactics, challenge the synthesis of learning, and explore fears around sharing the idea. He emphasized that even small amounts of cognitive offloading enabled more thoughtful, tailored guidance: {~\em``The more you can offload, the more you can connect the dots.''}

Both novices and the mentor reported that the system-supported coaching sessions were more focused, productive, and in-depth. P2 appreciated that the dashboard provided a {~\em``common understanding''} that saved time otherwise spent on updates. P8 noted the meeting {~\em``accelerated the topic to where [they] needed to go faster.''} The mentor also confirmed that {~\em``because the groundwork was already done, [they] could jump straight into more advanced work''}---e.g., reviewing prototypes or designing validation experiments, which would normally occur in later meetings. The mentor also emphasized that a clear, pre-articulated goal improved feedback quality and consistency: {~\em``It’s hard to give feedback when you don’t know where you’re trying to go.''} In contrast to previous meetings without structured agendas---where conflicting mentor advice sometimes confused students---the system fostered a concentrated effort on mutually prioritized risks.

Despite these benefits, some novices initially misunderstood the collaborative role of the system, expecting it to behave like a reactive AI assistant rather than a preparatory tool for human- human interaction. Several participants (P1, P8, P10) approached the system with the same mindset they brought to tools like ChatGPT, anticipating immediate answers rather than reflective preparation. P10, for instance, admitted that he initially found the system unhelpful, viewing it as just another tool meant to provide immediate answers. However, his opinions changed after the coaching meeting. He reflected, {\em``I can definitely see how it helps [the mentor] personalize his suggestions to me.''} He appreciated this combined approach, noting, 
\begin{quote}
    {\em``I like the approach of thinking about risk and being a complementary tool for a mentor... It just helps me think about things. It feels nice to not tell me what to do, but to make me aware of other things, and then let’s work with a human to make the journey more focused.''} 
\end{quote}
His perception shifted after the meeting once he realized that the system was intended to complement the mentor's role, acting as a tool in collaboration with the mentor rather than in isolation.

\section{Discussion}
The findings presented above demonstrate how our system supported proactive risk diagnosis, reflection, mentor preparation, and emotionally attuned interaction---across asymmetric roles in entrepreneurship coaching, a real-world, ill-defined problem domain. In this section, we reflect on these contributions in relation to prior work, outline design challenges and future opportunities, and discuss how the system’s design principles offer broader implications for CSCW, HCI, and human-AI collaboration.

\subsection{Moving Beyond Static Templates and Reactive AI: A New Model for Supporting Coaching in Ill-Defined Domains}

Much of the prior work in entrepreneurship coaching has relied on static templates and checklists to help novices structure their ideas or provide updates for mentors~\cite{reeslewisBuildingSupportTools2015, riesLeanStartupHow2017}. While valuable, these tools assume that novices possess advanced metacognitive skills to articulate and diagnose risks---an assumption that often does not hold in ill-defined domains like entrepreneurship~\cite{carlsonDesignRisksFramework2020}. Moreover, their rigidity fails to capture the evolving, uncertain nature of real-world ventures~\cite{macneilProbMapAutomaticallyConstructing2021}. Our system moves beyond these limitations by providing proactive and adaptive scaffolding that surfaces potential risks---rather than asking novices to self-diagnose---and then guides reflection through tailored, context-sensitive questions.  This reduces novices' cognitive burden while preserving agency, transforming preparation into a process of interactive diagnostic reasoning rather than static form-filling.

In contrast to Intelligent Tutoring Systems (ITS) that rely on predefined problems and solutions~\cite{vanlehnBehaviorTutoringSystems2006a, koedingerCognitiveTutorsTechnology2006}, our system operates in ill-defined spaces without requiring an exhaustive problem taxonomy. It leverages a cognitive coaching model built from expert mentoring practices~\cite{carlsonDesignRisksFramework2020,huangIntelligentCoachingSystems2023}, but crucially, this model is not static. Mentors can inspect, revise, and extend the logic that drives the system---modifying risk categories and phrasing as their needs evolve. This design supports both domain specificity and long-term adaptability, allowing the system to reflect the lived reasoning of expert coaches without encoding fixed rules.

LLM-powered systems like \textit{Thinking Assistants}~\cite{parkThinkingAssistantsLLMBased2024} and \textit{Rehearsal}~\cite{shaikhRehearsalSimulatingConflict2024} mark a shift toward reflective, theory-informed AI, but they remain limited in scope. Designed primarily for individual use or scripted simulations and assuming static expert knowledge, these systems are not built to support the relational, asynchronous, and co-constructed nature of mentorship. Our work fills this gap by treating AI as collaborative infrastructure---not simply assisting users but facilitating shared understanding, alignment, and strategic preparation across asymmetrical roles. This coordination enabled coaching sessions that were more focused, intentional, and in-depth---participants described meetings as more efficient, better aligned, and reaching greater depth earlier. Furthermore, our approach treats expert knowledge as a living, co-authored resource where mentors can shape the model without technical overhead. This approach extends the role of domain experts from informants to co-designers of intelligence. This is particularly critical in coaching, where expertise is relational, shaped by lived experience, and often difficult to formalize in advance. It also offers a new model for human-AI collaboration: one where experts don’t merely supervise the AI by accepting or rejecting AI outputs~\cite{patelHumanMachinePartnership2019, cranshawCalendarHelpDesigning2017, caiHelloAIUncovering2019} but actively govern, negotiate, and evolve its logic as part of their ongoing work. 

By combining proactive and adaptive risk diagnosis with expert-governed cognitive models, this work proposes a new paradigm for designing AI systems to support complex, relational human work. It reimagines AI not as a tool for completing tasks or generating answers but as a diagnostic scaffold and facilitator of judgment. This model replaces static scaffolding with dynamic alignment, allowing AI to adapt alongside users and support real-time sense-making across evolving, high-uncertainty domains. In doing so, it expands the design space for collaborative AI systems---from assistants that react to mediators that help humans reason, prepare, and collaborate more effectively.

\subsection{Challenges and Design Opportunities}
While our system succeeded in scaffolding critical thinking, preparation, and mentor-novice coordination, several design challenges emerged that offer directions for future research.  A major issue was misaligned expectations, particularly among novices accustomed to reactive tools like ChatGPT. Many expected immediate answers or prescriptive advice and were caught off guard when the system posed reflective, diagnostic questions instead. This mismatch undermined engagement and occasionally led to shallow or minimal input. It highlights a broader issue in HCI and CSCW: As AI becomes more present in everyday tools, user expectations of ~\textit{“what AI is for”} are increasingly shaped by mainstream models, which often prioritize speed, fluency, and directive output~\cite{sarkarAIShouldChallenge2024, jacobsDesigningAITrust2021c, lugerHavingReallyBad2016, zamfirescu-pereiraWhyJohnnyCan2023}.

One design idea is to guide users toward a new understanding of what ~\textit{``help''} looks like. Proactive AI systems, especially those designed for scaffolding metacognition or collaboration in ill-defined domains, must carefully choreograph their introduction and use. Onboarding should not merely explain what a system does---it should position the user within a different kind of relationship with the AI. For instance, early interactions might explicitly model the system’s questioning role or progressively reveal its diagnostic intent through layered examples~\cite[e.g.,][]{lugerHavingReallyBad2016}.

Despite its benefits, the system raises important questions about the risk of over-scaffolding novice thinking. Unlike static templates, which require novices to diagnose risks on their own, this system surfaces risks on their behalf and guides reflection through targeted prompts. While this lowers cognitive barriers and supports critical thinking, it also reduces the need for novices to engage in self-directed diagnostic reasoning---a skill that is central to entrepreneurial learning~\cite{reeslewisAssessingIterativePlanning2019, carlsonDesignRisksFramework2020}. Prior research in the Learning Sciences has cautioned that overly structured supports can hinder long-term metacognitive development by displacing the cognitive work of identifying problems and making sense of them~\cite{dillenbourgScriptingCSCLRisks2002, kirschnerWhyMinimalGuidance2006}.

Our system provided just-in-time scaffolding~\cite{quintanaScaffoldingDesignFramework2004} to highlight diagnostic blind spots, critical to guide novices toward more strategic, reflective discussions with their mentor. Given that the system did not give them conclusions, only surface diagnostic entry points, we are arguably still operating at a safe distance from over-scaffolding. Still, to guard against dependency over time, future iterations of the system should move toward graduated scaffolding~\cite{collinsCognitiveApprenticeshipTeaching1989, biswasLearningTeachingNew2005}, where the system prompts novices to take on more of this reasoning themselves over time. Another direction is to design AI systems that encourage novices to reflect not just on their initial risks but also on their diagnostic reasoning processes. This draws from the tradition of metacognitive reflection tools in HCI ~\cite[e.g.,][]{baumerReviewingReflectionUse2014}, which show that systems that scaffold reflection \textit{on thinking}---not just \textit{on content}---can foster transferable skills and long-term cognitive growth. For example, the system might occasionally ask, ~\textit{``How did you decide this was your most critical risk?''} or ~\textit{``What information might change your diagnosis?''}.

We propose that future proactive AI systems in ill-defined domains such as entrepreneurship should not only scaffold user thinking but also scaffold how users learn to scaffold their thinking. That is, the goal should not be static personalization but dynamic transfer: helping users internalize the diagnostic strategies embedded in the system so that they can eventually operate with greater independence. This approach aligns with research on teachable agents~\cite{biswasLearningTeachingNew2005} and reflective informatics~\cite{baumerReviewingReflectionUse2014} and presents an important opportunity for human-AI systems to foster both immediate utility and long-term skill development. Supporting this trajectory will likely require new design patterns, such as visualizing the growth in diagnostic complexity over time or prompting users to reflect on how their risk prioritization has evolved. In doing so, AI systems can act not just as intelligent assistants but as partners in developing human judgment.

\subsection{Implications for Other Domains}
While this study is focused on entrepreneurship coaching, the design principles we identified have broad applicability to other domains such as knowledge work, healthcare training, education, peer mentoring, and creative design. These domains are characterized by ill-defined problems and all share the core qualities with entrepreneurship coaching: uncertainty, emotional stakes, cognitive complexity, and the need for nuanced social support~\cite{jacobsDesigningAITrust2021c, kulkarniPeerStudioRapidPeer2015, reeslewisEncouragingEngineeringDesign2023, holsteinCoDesigningRealTimeClassroom2019}. These are precisely the conditions where proactive, context-aware, and relationally attuned AI systems may have the greatest impact.

For example, the principle of designing AI that proactively challenges rather than obeys has clear relevance in domains where untested assumptions or blind spots can undermine progress. In product design~\cite{reeslewisAssessingIterativePlanning2019}, civic deliberation~\cite{wuEngagingStakeholdersDeliberation2024, umbelinoIncreasingInclusionTimeEfficiency2023}, or scientific research~\cite{zhangAgileResearchStudios2017, gargOrchestrationScriptsSystem2023}, AI agents could play a proactive role by surfacing counterpoints or reframing questions that prompt users to reflect more deeply---much like a critical peer might in a design critique. Rather than reinforcing the user’s framing, AI can act as a {~\em``cognitive provocateur''}~\cite{sarkarAIShouldChallenge2024}, encouraging collaborative sensemaking and interpretive flexibility~\cite{gmeinerExploringChallengesOpportunities2023, cranshawCalendarHelpDesigning2017}. Similarly, in educational contexts, prompting learners to question their assumptions---rather than offering corrective feedback---has been shown to promote metacognitive development~\cite{zhangAgileResearchStudios2017, carlsonDesignRisksFramework2020}.

The principle of layering dual context to adapt support also generalizes across domains like education, healthcare, and even remote team collaboration, where people’s goals, challenges, and capacities are different and change over time~\cite{jacobsDesigningAITrust2021c, gargOrchestrationScriptsSystem2023, reeslewisBuildingSupportTools2015}. Systems that dynamically adapt scaffolds based on the progression of the learner~\cite{holsteinCoDesigningRealTimeClassroom2019}, the emotional states of both the patient and the doctor~\cite{mentisConcealmentEmotionEmergency2013}, or the evolving status of the project~\cite{gargOrchestrationScriptsSystem2023} are better suited to maintain meaningful support. Such systems must treat context as a living structure, continuously updating their model of the user’s situation to personalize both what support is offered and how it is delivered~\cite{gargOrchestrationScriptsSystem2023, holsteinCoDesigningRealTimeClassroom2019,mentisConcealmentEmotionEmergency2013, dourishWhatWeTalk2004}.

The principle of empowering mentors to govern AI logic addresses a growing need across domains like clinical decision-making, education, and scientific modeling. In these settings, experts often seek not just assistance but systems they can trust, critique, and co-adapt with~\cite{jacobsDesigningAITrust2021c}. When mentors or domain experts can inspect how an AI diagnoses a risk or chooses a prompt---and modify that reasoning if needed---they maintain agency and avoid becoming passive overseers of a black box~\cite{amershiGuidelinesHumanAIInteraction2019, yangReexaminingWhetherWhy2020a}. For example, a teacher might want to adjust how a tutoring system interprets the ~\textit{``risk of disengagement''} or a physician might want to revise what counts as a red flag for follow-up care.

Finally, the principles of surfacing root causes like emotional risks and scaffolding human- human interaction extend to any domain involving social, emotional, or relational labor---such as peer mentorship~\cite{kulkarniPeerStudioRapidPeer2015}, healthcare supervision~\cite{jacobsDesigningAITrust2021c}, design feedback~\cite{reeslewisBuildingSupportTools2015}, or even civic facilitation~\cite{wuEngagingStakeholdersDeliberation2024, umbelinoIncreasingInclusionTimeEfficiency2023}. In many of these settings, emotional hesitations (e.g., fear of failure, perfectionism, avoidance) are deeply entangled with strategic action yet remain invisible. When AI systems help make those risks legible---without pathologizing or overstepping---they enable more compassionate, focused, and effective collaboration~\cite{jacobsDesigningAITrust2021c, holsteinCoDesigningRealTimeClassroom2019}. Equally important is how these systems prepare both parties---mentor and mentees, supervisor and team member---with shared reflective context before a conversation, priming richer and more efficient human dialogue~\cite{clarkGroundingCommunication1991}.

Taken together, these design principles offer a framework for designing AI systems that do more than supporting individuals on task automation---they support human relationships, shared reasoning, and reflective learning in settings where outcomes emerge through dialogue, not automation. By exploring how these principles play out across different domains, future research can help develop AI systems that not only enhance productivity but deepen our capacity to understand, support, and challenge each other more intelligently and humanely.

\section{Limitations and Future Work}
A key limitation of this study lies in its scale and scope. The deployment involved a mentor and eleven novices, each using the system in a single coaching meeting. Although this limits generalizability and long-term insights, the focus on a single mentor was intentional, given the deep integration of the system with a cognitive coaching model. Customizing the system to one mentoring style allowed us to investigate how AI could scaffold complex, ill-defined coaching conversations without introducing confounding factors from mentor variability. In line with the RtD methodology~\cite{zimmermanResearchDesignHCI2014}, the study prioritized depth over breadth, using situated design and in-the-wild evaluation to generate transferable insights on proactive AI support in real-world mentoring.

This narrow scope also limited our ability to examine longitudinal effects, such as trust dynamics, evolving diagnostic skills, or shifting mentor-novice relationships. Our immediate steps for future work will extend the deployment across multiple mentors and sessions to explore how the system scales across coaching styles, venture development stages, and team dynamics. Longer-term studies would also allow for richer reflection on how users internalize AI scaffolding strategies over time. This expansion aligns with iterative RtD approaches, where early design interventions seed future refinements, adaptation, and theory building in diverse contexts.

The university incubator context provided a unique opportunity to study proactive AI systems in a resource-rich and high-agency environment. The participants were fluent in digital communication, highly motivated, and already engaged in reflective coaching. Although this enabled clean observations of how AI could scaffold strategic thinking and emotional insight, it also limits the applicability of findings to less-resourced settings. As scholars have argued~\cite{costanza-chockDesignJusticeCommunityLed2020, toyamaTechnologyAmplifierInternational2011}, systems developed around privileged users can reproduce inequities if applied broadly without adaptation. Future work should extend this design to more diverse communities through co-design, multi-modal interaction, and scaffolding strategies that promote access, equity, and transferability across educational and technological divides.

\section{Conclusion}
In this paper, we presented a proactive human-AI system that supports entrepreneurship coaching by surfacing risks, scaffolding novice reflection, and enabling mentors to inspect and adapt system logic. Through a field deployment, we examined how the system supported meta-cognition, emotional insight, and more intentional mentor–novice collaboration. This work contributes a new system architecture for human-AI collaboration that integrates cognitive coaching models with LLMs to support domain-specific reasoning; empirical evidence for CSCW on how AI can augment, not replace, human-human collaboration; and design principles on scaffolding meta-cognitive skills like diagnosis and strategic planning in ill-defined domains. We envision future systems that extend these ideas across domains where AI can serve as infrastructure for meaningful, adaptive, and reflective human---human interaction.

\bibliographystyle{ACM-Reference-Format}
\bibliography{new}

\appendix
\section{Appendix}
To support context-aware, expert-informed reasoning, our system uses a modular prompt chaining architecture composed of four distinct types of prompt templates. Each prompt is designed to complete a targeted reasoning task and is informed by structured domain knowledge and novice-specific project contexts stored in a real-time database. The outputs of earlier prompts feed into subsequent prompts, forming a dynamic pipeline (see Figure~\ref{fig:chained_prompts}).

Below, we detail the role and design of each prompt template. To aid reproducibility, a complete example of a chained prompt sequence is available in the supplementary materials.

\begin{itemize}
    \item \textbf{Project Context Tagging Prompt}: Extracts structured representations from novice input into predefined schema fields (e.g., assumptions, plans, target users) based on a project model~\cite{huangIntelligentCoachingSystems2023}. This enables downstream prompts to work with machine-readable key-value pairs instead of raw free text. To minimize hallucination, the LLM is explicitly instructed to extract only verifiable statements and return a JSON object, improving both auditability and consistency.

    \item \textbf{Risk Diagnosis Prompt}:  Evaluates novice project data against predefined risk categories from the coaching model~\cite{carlsonDesignRisksFramework2020}. Given project contexts collected through novices' inputs, it identifies one or more relevant risks (e.g., vague value proposition, no validation plan), justifies each diagnosis with reference to input data, and returns structured output (risk name and rationale). The prompt includes risk definitions and examples, aligning with best practices in few-shot prompting and model steering~\cite{zamfirescu-pereiraWhyJohnnyCan2023}.

    \item \textbf{Question Generation Prompt}: Formulates questions on predefined project areas and tailored to novice-provided contexts. When given a diagnosed risk and its justification, it also generates speculative and open-ended follow-up questions to help novices reflect on the underlying issue. Inspired by work in affective and reflective prompt design~\cite{xuJamplateExploringLLMEnhanced2024, wangBridgingNoviceExpertGap2024}, the prompt instructs the LLM to adopt a tone that is supportive and curious, encouraging engagement while avoiding prescriptive or judgmental phrasing.

    \item \textbf{Coaching Strategy Suggestion Prompt}: Tailored for mentor use, this prompt synthesizes risks, novice goals, and mentor-specified coaching priorities into strategy suggestions. These suggestions include targeted coaching questions and hypotheses about potential root causes (e.g., fear of failure, lack of clarity). The prompt incorporates mentors' specified coaching goals to ensure that the suggestions are practical, nuanced, and tailored to those goals. 
\end{itemize}

\begin{table}[ht]
\small
\centering
\resizebox{\textwidth}{!}{%
\begin{tabular}{|p{3cm}|p{6cm}|p{6cm}|}
\hline
\rowcolor{lightgray} 
\textbf{Project Areas} & \textbf{Description} & \textbf{Example Questions} \\
\hline
Project information & The overview of a novice’s venture, including information about the problem this venture aims to solve, and the proposed solution to solve that problem. & What is the problem you are trying to solve, and what is your proposed solution to solve this problem? \\
\hline
Current Focus & The specific aspect of the venture that the novice is currently focusing on and taking action on. & What specific aspects of your venture are you currently focusing on? What actions are you taking to make progress on that? \\
\hline
Learning & The most useful and critical learning that the novice has gained recently about their venture. & Is there any learning that has been particularly beneficial or critical for your venture? \\
\hline
Obstacles & Obstacles or roadblocks that are slowing the novice down. & Is there anything that is currently slowing you down? \\
\hline
Plan & Goals that the novice plans to accomplish in the next few weeks. & What goals are you planning to accomplish in the next few weeks? \\
\hline
Coaching outcome & Specific outcome that the novice is looking to achieve through the next meeting with the mentor. & Looking ahead, what is a success metric that will make your next coaching meeting worthwhile? \\
\hline
Emotions & Emotions that the novice is currently experiencing with their project. & How would you describe your feelings lately? Excited? Nervous? \\
\hline

\end{tabular}
}
\caption{The project model outlines key areas such as problem, practice, and desired outcome that novices should articulate before coaching meetings and includes example questions to probe novices.}
\label{tab:s3-projquestions}
\end{table}

\begin{table}[ht]
\small
\centering
\resizebox{\textwidth}{!}{%
\begin{tabular}{|p{4cm}|p{11cm}|}
\hline
\rowcolor{lightgray}
\textbf{Risk} & \textbf{Description} \\
\hline
Communicate with customers & If novices do not clearly articulate and communicate their brand promise and how the product delivers on it, there is a risk that customers may perceive the solution as inadequate. \\
\hline
Customers and needs & If novices cannot articulate customers' needs that are supported by evidence, there is a risk they will misconstrue the root cause(s) of that need and design ineffective solutions. \\
\hline
Distribution channels & If novices do not know how they will distribute the solutions or if they lack evidence that their strategy will work, there is a risk of designing something that never goes into customers' hands. \\
\hline
Existing solutions & If novices have not thoroughly researched existing solutions, and cannot articulate why their solution is superior to those existing solutions, there is a risk that the customer will not adopt it. \\
\hline
Identify risky assumptions & If novices have not identified and validated risky assumptions in their ideas and concepts, there is a risk of these unvalidated assumptions hindering their company's growth and adoption. \\
\hline
Perfectionism & If novices have built a product but have been delaying showing the product to customers, there is a risk that they are being perfectionist. \\
\hline
Planning & If novices' goals are not actionable, feasible, and measurable, or based on important risks, there is a risk that they may end up doing busy work that does not produce value nor help them progress. \\
\hline
Raising capital & If novices are overly focused on raising venture capital, there is a risk that raising money is the trophy they seek at the expense of building a great product and business. \\
\hline
Teamwork & If there is a lack of cohesion or alignment on buy-ins and expectations among team members, there is a risk for teamwork to negatively affect the venture's progress. \\
\hline
Testing & When novices test their products, if they do not have valid processes and measurable, specific metrics for success, there is a risk of not making progress toward a solution the customers want. \\
\hline
Value propositions & If novices cannot explain and provide evidence of how their solution will solve the customer's problem, there is a risk that it will not. \\
\hline
\end{tabular}
}
\caption{The risk model outlines and articulates common risks that novices encounter, such as not understanding customers' needs and not having evidence on how the proposed solution will solve customers' needs.}
\label{tab:s3-riskmodel}
\end{table}

\end{document}